# The Eyes Have It!: Using Human-Selected Features for Predicting Athletes' Performance

Jaeyoun You

you.jae@snu.ac.kr, Department of Intelligence and Information, Seoul National University

Jinhan Choi

jinhanchoi@snu.ac.kr, Department of Intelligence and Information, Seoul National University

Ho-Jae Shin

fromnight0@ajou.ac.kr, Dasan University College, Ajou University

Bongwon Suh

bongwon@snu.ac.kr, Department of Intelligence and Information, Seoul National University

Predicting athletes' performance has relied mostly on statistical data. Besides the traditional data, various types of data, including video, have become available. However, it is challenging to use them for deep learning, especially when the size of the athletes' dataset is small. This research proposes a feature-selection strategy based on the criteria used by insightful people, which could improve ML performance. Our ML model employs features selected by people who correctly evaluated the athletes' future performance. We tested out a strategy to predict the LPGA players' next day performance using their interview video. We asked study participants to predict the players' next day score after watching the interviews and asked why. Using combined features of the facial landmarks' movements, derived from the participants, and meta-data showed a better F1-score than using each feature separately. This study suggests that the human-in-the-loop model could improve algorithms' performance with small-dataset.

# 1 Introduction

In view of the belated Olympics approaching, athletes are trying to boost their skills. In addition to physical training, statistical analysis is employed to obtain better performance. As the sports data is well-refined and well-stacked, they are determined to be suitable for machine learning [8]. For example, in baseball, with the increasing use of the on-base percentage (OBP), the sabermetrics shifted the ballgame strategy paradigm into a form of aggressive use of data [47]. In a horse-racing score prediction run by back-propagation model reached to 77% accuracy [12]. Researchers in that study used structured data such as horse weight, racing distance, trainers, jockey, and track conditions. Computer vision-based predictions, using unstructured data, are also increasingly utilized in sports. There exist models to foresee the direction of the cricket ball in real time [66], and predict swing performance (fade, hook, straight shot) through pose estimation of the golfers [38]. Based on stacked data, with accurate features and algorithms, deep learning could now contribute to sports analysis.

Algorithm researches focusing on an athlete's inner side have also advanced steadily. Primarily, many studies have used brain signals, heart rate, and blood pressure. One study estimated the anxiety of a player by measuring heart rate and revealed that it drives better motivation for putt [72]. A study has shown that professional golfers' cerebellum is better enhanced in functional connectivity; thus enabling them to produce better shots with constant swing speeds and backswing angles than amateur golfers [39]. As psychological features, automaticity and commitment to golf could bring nice strokes, as reported in other sports research [28].

Previous studies were limited to a posteriori analysis. Some attempts have been made to predict performance with near real-time data. In the U.S. Tennis Open, IBM recently worked on a program of real-time analysis with recent scores and articles [32]. However, their predictions are based on perspectives of others rather than the athletes. There exists a study that involves interviews immediately after the golf tournament to measure fresh minds that ends up with good results; however, the number of players involved was limited [65]. The immediate analysis of the athlete condition is significantly valuable for the athlete herself/himself and the coaches. Their physical and psychological states may vary slightly every day. Early grasp leads to a better strategy to obtain the best performance.

However, in representation learning, we need a large amount of personal data for athletes. A crucial issue of in-situ result analysis remains a chance to be discovered. However, we cannot make athletes do self-reports before the game, and cannot estimate dopamine in a real-time match. On the other hand, real-time human data that does not prove bothersome or disturbing to athletes will soon be an issue for sports analysis. In addition, we need to overcome the small-dataset issue. Therefore, we introduced the knowledge of insightful people, having good judgment on others' psychological state, into our algorithm, and investigated its contribution to a better prediction.

This paper outlines a new approach to sports analysis, proposing a feature-selection strategy based on the criteria used by insightful people, which could improve the algorithm performance. In our knowledge, performance predictions of athletes are mostly based on minimal data. Our model employs features selected by people who correctly estimated future performance of athletes after watching their interview videos. We used features of the facial landmark movements derived from the participants and meta-data of athletes. As a result, using combined features demonstrated a better performance than using each feature separately. This paper sheds new light on how the human-in-the-loop model could improve algorithm performance with a small dataset, in the case of performance predictions in sports. In the following sections, we will discuss the research area in which human computer interaction (HCI) could contribute.

# 2  RELATED WORK

In this research, we propose a strategy based on people to improve machine learning performance. For evaluation, we employed a sports score prediction task using interview videos of athletes. Therefore, we needed to investigate the following studies in advance. First, we considered the relationships between athletes' psychological state and performance. In studies of sports psychology, these links are the main arguments. We explored recent sports studies in HCI in this part. Second, we reviewed two different approaches to identify human emotions by reading faces. Lately, the dominant discourse of somatic markers has been challenged by counter-arguments based on empirical studies. In this *status quo*, we expected to find some supplement to emotion recognition models. Third, the deep-learning analysis of sequential data, such as micro-expressions, was studied. By reviewing these flows, the collaboration between machine perception and human insight was discussed as a solution to the small dataset learning issue of athletes' emotion analysis.

## 2.1 Sports Psychology

Sports psychology is the study of how psychological features affect athletes' performance [71]. Researchers in this area intensively check the athlete's personality and emotions. In sports, the definition of personality is summed up with the athlete's uniqueness, consistency, and tendency [26][31][22]. When we say, "He tends to lose his temper," "She has a fiery temper," we are talking about his/her personality, which is inborn himself/herself. Personality is quite personal and private; therefore, we should accumulate objectively evaluated temper data for each person to perform deep-learning analysis, which is a tricky task.

On the contrary, emotions have been studied based on the universality of expressions and occurrence, inspired by Ekman [16]. Mainly, athletes' anxiety related to performance has been studied extensively. Researchers have measured the anxious state using EEG, heart rate, blood pressure, and pattern of facial muscle movements [24][50][61]. Lately, the model of individual zone of optimal functioning (IZOF), which functions on the principle that anxiety levels could vary according to the individual, is highlighted [27]. In the anxiety scale from zero to one-hundred, athlete A showed the best ability on 40 when other players needed 60 to the scale. Because the links between emotions and performances vary among individuals [34], we need accumulated personal data again.

There are two ways to measure an athletes' emotions: first, coaches watching the athletes' behavior and filling in a checklist; second, the athlete performs a self-report after the match. Especially for the latter case, athletes undergo 'competition reflection', which reports their thoughts, attention, concentration, motives, and circumstances during the event. These features contribute to building a psychological strategy for the best performance play [57]. Lazarus, an authority on stress studies, also deployed these data to build the cognitive-motivational-relational (CMR) theory of emotion, which is an emotion occurrence chain. The framework considers how athletes perceive relationships and make motivations; in other words, it considers which emotions emerge [43]. Merely, research in sports psychology has tended to focus on emotions' *a priori* process. We have focused our attention to a more integrated analysis.

In the literature on CHI, there exist several accomplishments regarding the interaction between athletes and machines. In CHI 2014, the workshop named 'HCI and Sports' suggested four areas where HCI could contribute to sports: a bodily control and awareness, sports motivation and fun, pain and discomfort, and current use and UX of existing sports [54]. In the last two years in CHI, more data-driven, immersive experiencing research has shown up; the players' behavior analysis in e-sports [62][60][41]; growing interaction for regular exercise [3][63][20]; VR experience for the immersive sports experience [35]; informatic swimming goggles for improving directional perception [40]; data-driven approaches for the audience [37][77][55]; and the interaction study of facial electrical stimulation contributes to helping professional athletes' training [53]. Much work on the role of HCI in sports analysis has been carried out, yet there exist few studies on the athletes' psychological state. We investigate the areas where HCI could contribute to sport psychology.

## 2.2 Emotions

The facial action coding system (FACS) model by Ekman now became the basis of modern emotion recognition systems and affective computing theories. This model relates basic emotions, including anger, surprise, sadness, happiness, disgust, and fear, to facial expression patterns derived from a vast facial dataset [14][17]. The underlying hypothesis of the model is the theory of 'somatic markers' by Damasio. According to his hypothesis, the human body and emotion are under the system of the human brain. If external stimuli occur, bodily changes respond very quickly because of the 'as-if body loop' mechanism. The loop, evolved in human beings, makes the vital system to maintain homeostasis. Thus, all the responses, including emotions, are performed naturally and universally identical because they are the outcome of evolution [10].

The social-constructionist view, leading by Barrett, on the contrary, claims that the emotions are the product of individuality. It raised objections, arguing that universality, in which somatic markers and FACS guarantees have exceptionalities, depends on cultural backgrounds or individual contexts. According to the constructive assertion, emotion is not localized in any brain area; thus, it cannot appear in the same expression. It also refutes the six-basic emotion. According to constructionism, adjectives expressing emotions may vary depending on which language people use, so the emotion layer can also be distinguished [30][6][33]. Therefore, theorists warned that emotion recognition algorithms structured by cultural bias could lead to technical misuses. Ekman criticized that the claim was based on insufficient evidence [19].

We believe both sides clash because they argue within causality and correlation from each of their perspective. Damasio testified that sadness was intensified by impairment in the specific brain part in medical grounds [10]. However, constructionists criticized that the sadness occurrence is not based on the spot being damaged; they stated that the brain change only modified the body system and did not affected the sadness system directly [2]. According to medical opinion, the brain area is a sufficient condition for the occurrence of sadness. Inversely, that spot damage is not the necessary condition for sadness, as constructionists argue. To the best of our knowledge, we cannot deny that the brain area correlates with sadness. If both theorists accept a 'correlative' rather than a 'causational' view, about the brain spot where emotion is affected, we expect that they could find an agreement point to emotion theory.

In addition, both theories relate to the occurrence of emotions. Their approaches are not well suited to the emotion reader's perspectives. For reading emotions, Damasio only explains sympathy as the inherent evolutionary system being stimulated by external factors. Constructionists gave a brief account that people read others' minds depending on their personal experience and contextual information. In phenomenology, researchers have tried to explain with intersubjectivity; they proposed two conditions for intersubjectivity: the isomorphism, that one's mind is the same as others', and the empathetic intuition, that knowing others' states directly without mediation [68].

Damasio's theory and constructionism both cannot satisfy those qualifications, so that phenomenological insights can open us the door to new research possibilities that have not been tried so far. Phenomenology of intersubjectivity claims that we can directly perceive the minds of others, not only by relying on primary factual data from their bodies, but also by using some contextual frameworks for interpretation which are based on the experiences of life shared by us and them. Therefore, in this research, we took a step back from the rigid alternative between classical or constructionist theories of emotions, and reconsidered reading other's minds by borrowing a phenomenological insight and approach to some extent. By perceiving others' changes in face or gestures, we linked them to her/his meta-data, and then produced specific patterns that only the machine could understand. We used this algorithmic pattern as an alternative to judge others' states.

## 2.3    Emotion Recognition in Computer Vision

Micro-expressions, such as an fleeting look in the face, is an active research area in affective computing. Ekman defined micro-expressions as hidden emotions derived by the neural system and expressed in 0.25 seconds [18]. He explained it as "involuntary emotional leakage," which exposes a person's genuine emotions [15].

Therefore, in a micro-expression study, detection of the fraction of a second in the sequence video is crucial. Researchers commonly split the video into pixel-levels and identify the detailed movements of facial muscles [73]. In another study, researchers gathered diverse micro-expression data in various cultural backgrounds and adjusted thresholds on neutral expression [11]. This seems to be an innovative approach because they measured the default mode, leading to bias without a cultural consideration. However, there is still considerable uncertainty concerning the FACS model, in which researchers built their hypothesis.

A recent systematic study on automatically recognizing spots for micro-expressions in video was carried out by Li et al [48]. Researchers measured quick changes in facial expressions and automatically detected peak points to find positive and negative emotions and surprise. However, the study has a limitation that blinking eyes or nodding heads generate negative effects. To overcome the disadvantage of natural facial behavior, we selected several facial landmarks, such as eyes and lips, and did not use all the facial muscles. Instead of leaving all the decision-making process to the machine, we actively employ human insights for better estimation.

An increasing number of studies have found that micro-expressions could be practically used in human-computer interaction. To automatically diagnose depression, researchers have analyzed the dynamics of facial expressions [79]. In AI speaker studies, researchers make speakers give proper feedback to users, while the camera detects frowning faces from users who are given unwanted answers from the machine [74]. Babaei et al. also worked on facial expression changes to investigate attentional state [5]. With a camera on the office monitor, they tracked fifteen participants' keyboards-tapping and mouse-clicking behaviors combined with their facial muscle transitions. This work highlighted faces as a potential indicator for attention detection. Our work will further explore a combination of facial cues and personal data to analyze the specific psychological state.

The micro-expression mentioned by Ekman was originally used for detecting lies and discovering unusual states of individuals. For high accuracy, personal characteristics should be gathered to compare, as the researcher mentioned [15]. While computer vision studies require a large quantity of random data, repeated data belonging to the same person could hardly be collected. Thus, emotion labels used in face recognition are based on the expressions of majorities. L. Herbert, who criticized the use of FACS at the U.S. airport in the 9/11 terrorism, claimed that 'Facism (face+racism)' could have occurred by face recognition tools [29]. The technology itself embodied bias on minorities could lead Othello's errors to the authorities. Even if she/he is telling the truth, minorities could appear to be lying because of high tensions in the system trained by majorities. We strongly agree with this opinion; therefore, we will reflect individuals' diversity into the algorithm.

## 2.4 A Collaboration between AI and Human Vision

The machine perceives visuals with patterns based on pixels [44]. It has the strength to identify patterns invisible to humans, such as detecting the optical flow of ice [51]. With the prediction of the action or the situation in the near future, computer vision could contribute to giving warning signals in advance. Interestingly, there exists a challenge in predicting even unexpected scenes of the following movement with machine learning, with the Oops! dataset [21].

For guessing tasks, which are too vague to hit the mark, several papers argued that machine learning could contribute to the decision-making process. In the research on distinguishing artists' abstractions from kids' or animals' painting, machines demonstrated slightly higher accuracy than humans. Furthermore, the program analyzed where humans judge artisticity on images [64]. The study revealed that machines could detect instinctive senses in humans.

However, we need a massive dataset for such representation learning [36]. For example, for learning videos, Youtube-8M is essential [1]. Even the CIFAR-10 [42], which has sixty-thousands of images in ten classes, is thought to be too small for a deep convolutional neural model. Therefore, the modified VGG-16 model was developed to overcome the overfitting issue with this small dataset [49]. Recently, there has been an increasing number of few-shot learning challenges, handling tiny-datasets [78][70]. Even the fewshot-cifar100 or the mini-ImageNet, generally used in few-shot learning [56], are much larger than our research dataset. In early computer-vision studies, feature extraction was performed by human experts. We tapped back into past ideas.

In this context, research on human-in-the-loop recently highlighted the role of humans in the machine learning process. Like our challenge, there exist numerous tasks in which machine-alone cannot solve the problem. A rising number of studies discussing human ground theory on data could help human-AI collaboration [69][46]. In the task of remembering the detailed information from quickly passed scenes in the video, the vision algorithm combined with retrieved human information yielded better performance than other algorithms [45]. From this perspective, we devise the design of humans' ability to collaborate with machines as a method.

## 3 Concept of the study

In this section, we describe our theoretical framework on which our experimental design is based. As shown in Fig 1, the quadrant is about different approaches to recognize other's mind. The axis represents human versus machine, as well as classical theorists, including Damasio, versus Constructionism, including Barrett.

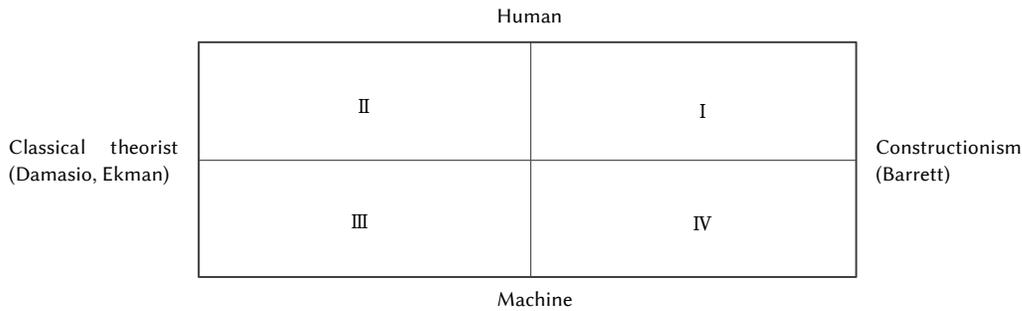

**Figure 1: Four different approaches to knowing other's mind is described as a quadrant. This framework will be the theoretical ground for our experimental design and methods.**

The first quadrant means, while judging others, humans use meta-information, such as sex, nationality, and hometown. These data are empirically built up. In this first quadrant(I), facial expressions are not included. In the second quadrant(II), human judges only with others' faces, with grounds constructed in society. For example, if someone frowns his nose, then the human judges him 'angry.' In the third quadrant(III), The machine reads faces with facial landmarks and match with pre-labeled emotions, such as happiness, surprise, and disgust. The fourth quadrant(IV) means machines use meta-data in their decision-making process. With structured data, the statistical analysis would fit in the fourth quadrant.

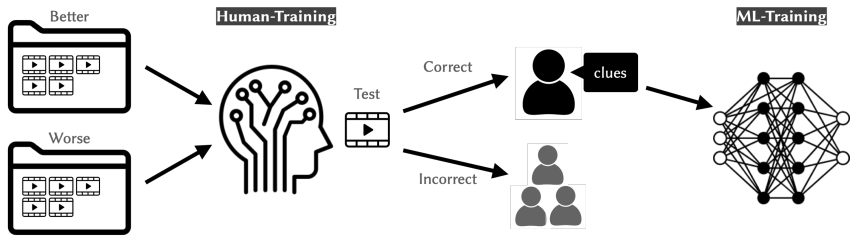

**Figure 2: The process of this research based on the theoretical framework described in Section 3.**

In this research, we came across all quadrants and adopted the four approaches described above, with the idea of phenomenology of intersubjectivity. We will use human perceptions based on their external information and viewpoints on others' faces. Further, we will accept the machine's perception of using meta-data and facial landmark detection. By combining different approaches, we will develop a basic framework for the loop mode for humans and machines, in which each of them can participate actively.

## 4  Study 1: cues from human insight

In this study, we explored human selected features which derived from the first(I) and the second(II) quadrant above. We use professional golfers' interview video for prediction modeling. Compared to previous methods using statistics, we focus on vision data and employ deep learning. However, a small amount of individual data is not sufficient for modeling machine learning algorithms. To come up with this problem, we planned to obtain cues from humans' guess behavior and apply them to algorithms. In this section, we determine what people selected for the prediction.

## 4.1 Data

We explored the Ladies Professional Golf Association (LPGA) tour data for the study. We gathered each round's interview videos and golfers' meta-data, including previous tour scores, nationality, age, and length of career. We chose golf because players need to keep positive flow states on themselves for four days in four rounds. Generally, the round-interview takes place immediately after the performance of that day, and mostly includes players who scored well in the round. We used interview videos to gain an account of the emotion from the player's immediate talk. We expected to measure the correlation between an emotion and a score, in the context of the word of Jim Flick [23], one of the gurus in golf; "The game of golf is 90% mental and 10% physical."

A total of 213 videos were used in Studies 1 and 2, which were filmed during 29 matches held between March 2013 and November 2018. As a major tournament, the Kraft Nabisco Championship (2013) and the Evian Championship (2016) were included in the dataset. The number of golfers was 74, from 18 nations such as the U.S., Republic of Korea, France, etc. We included various nationalities and ethnic groups, like other HCI researches focusing on the diversity of e-sports players [58]. On average, they took an interview at the age of 25.5, and 5.8 years from their turn to professional. The average height was 5 ft 7 inches.

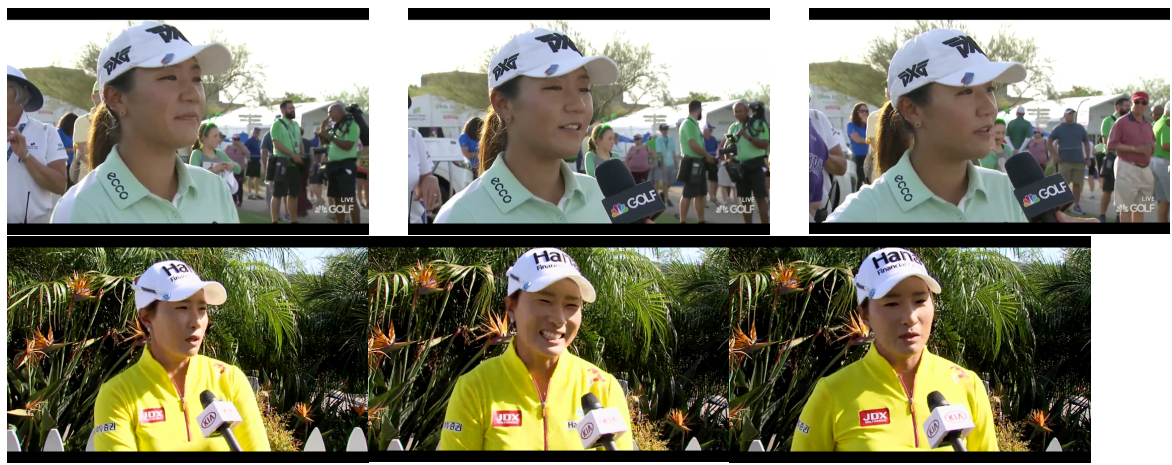

**Figure 3: Example of players' interview videos. copyright: LPGA, https://lpga.com/video/interview)**

We collected a total of 1476 on-site interviews from the official LPGA website (lpga.com) in advance. Subsequently, we chose more than 200 suitable clips under the following conditions: First, the clip requires the performance of the current day, together with the score of the next day. We excluded interviews conducted before the first round and after the final round, which did not fit the analysis. Second, we excluded videos on which python-based OpenCV library failed to extract faces, because the machine must detect and recognize faces delicately to model automated video analysis. The average running time was two minutes.

To analyze the correlation of the player's emotion and performance, we calculated a relative score for a dependent variable; as golfers do not play the same match for each round, and tournament organizers manage the handicap of each day by moving holes and pins, as well as the weather effects on absolute scores. To determine whether the player objectively did well or not, we compared the scores of other players. We calculated the proportion of the interviewee's next round's strokes by the average score and measured it against the interview day. A score that rose above or equaled it was labeled as 1, while the opposite was labeled as 0. We assigned a label to each score reflecting its relative score.

Meanwhile, we excluded audio during this research. In our ground truth, the content of the interview makes no difference. We considered the language skills, tones, and pitches of sounds could introduce bias.

## 4.2 Experiment and In-depth Interview

The aim of Study 1 was to investigate people's insight into others. Thus, we conducted a guessing test and in-depth interview for gathering grounds. First, we tested the participants for satisfying insightfulness. Then, we asked how they picked correct answers. We recruited totally twenty participants (including 5 females) aged between 26-55 years. Participants begin by sitting in front of a table with the 13-inch MacBook. We asked them to self-study with two different datasets then watch two videos to predict the athlete's future score. We allowed participants to use enough time to decide. On average, participants trained for 10 minutes and decided in 3 minutes each. The study took about half an hour, including an interview. Participants received a $5 coffee coupon as gratuity.

The detailed experimental design is as follows. Like the machine learns with a labeled training set, participants study with two data folders, labeled with 1 and 0, as described in Section 4.1. Each folder has five videos. To support participants' direct comparison, we showed videos in which the same player scored different outputs. For example, folder labeled-1 has golfer A's video resulting in a better score, and folder labeled-0 has golfer A's video resulting in a worse score. We chose five players with a diversity of their cultural backgrounds.

After learning plenty of time, participants watched two videos that were not included in the training set. Participants needed to answer whether players would play well or not on the next day. Subsequently, we asked the decision's ground from four options: golfer's facial expression, golfer's gesture, circumstances in the video, and participant's knowledge of golf. As soon as these steps have been carried out, we conducted an open-ended, in-depth interview.

## 4.3 Results

Only twenty percent of the participants hit the mark on two videos. Including participants who correctly answered at least one video, followed by the human accuracy rate increased to fifty percent. We considered them as insightful people and analyzed their answers. Because of a small number of participants and quizzes, there could be a criticism of the expression of 'insightful people.' We will discuss this issue in Section 6.2.

### 4.3.1 Demographic Information of Participants

First, we discovered participants' job, age, knowledge of golf, and confidence in their answers to identify their understanding of reading others.

As a result, knowledge of golf was found to be not crucial to guessing the next day's score. Moreover, the answerer's confidence did not match the correctness. Additionally, age did not affect the guess test. However, we could witness that individuals with some jobs, such as that of an instructor, tend to answer correctly.

**Table 1: Demographic Information of Participants**

| group | Jobs | Ages | Sex (Female) | Knowledge on Golf (0~5) | Certainty on Answers (%) |
|---|---|---|---|---|---|
| 2/2 corrects | Exhibition manager, Speech instructor, Oral-test instructor, PR team manager (previous journalist) | 30-40: 3<br>40-50: 1 | 1 | 2.25 | 67.5 |
| 1/2 correct | CEO, Office worker, Journalist, Graduate student | 30-40: 3<br>40-50: 2<br>50- : 1 | 0 | 3.66 | 65.83 |
| 0/2 correct | Public officer, Curator, Teacher, Graduate student, Developer, Lawyer, Office worker | 20-30: 2<br>30-40: 6<br>40-50: 2 | 4 | 2.2 | 71.6 |
| total | | 20 | 5 | 2.65 | 69.05 |

*4.3.2 Discourse within participants*

In the in-depth interview with open-end questions, the noun that correct answerers used the most is 'eyes (7 people)', followed by the 'gesture (5)' and 'confidence (5)'. Meanwhile, the 'lip (3)' and 'fatigue (3)' have also been mentioned multiple times.

As we could interpret as emotion, the confidence was generally referred to as 'facial expressions'. For example, P9 said that "(She looked like) she could not hide to surge up of her confidence." P4 told us that he was concerned about distinguishing 'pride' and 'excited state.' He said, "Excited-state ends up the bad score, but it is hard to tell apart from pride." Fatigue was also mentioned with eyes and gestures. "(those who will get less score on the next day) have teary eyes in common. I mean, I felt she wanted to go back to the room and took a shower(P2)".

In the facial landmarks, eyes were expressed with 'gaze,' 'attention,' and 'pupil's movement.' P13 mentioned, "We need to read their eyes (to know others' state)," and P19 said, "I worked 25 years for CEO, so I am quite skilled in behavior analysis. Even if she looks calm or happy, you can read her eyes inordinately quaver. I can feel that she is less sure of her score". Furthermore, P4 said, "(Those who will get less score on the next day) have trembling eye sights… I mean, she looked like she wants to leave the golf course as soon as possible."

Followed by eyes, lip is mostly mentioned in facial landmarks. P7 said, "(because there is no audio), I read speed through the shape of lips." Both P9 and P16, who received 100% correct answers, primarily focused on lips. P9 remarked, "golfer who would get a bad score on the next day takes more time on talking, and while they listen to the interviewer, they used to keep their mouth opened." P16 also stressed, "(for the golfer who will get less score on the next day), I read exhausted image on her… her lips motioned very fast."

Other words mentioned are 'gesture', 'sunshine', and 'score'. P20 mentioned gesture as a "passive attitude". P13 said, "Her facial expression was slightly obscure to figure out, so I refer to the scoreboard on screen. She shot nine-under on that day, so I thought she would not make such a perfect score on the next day, too." P7 was curious about the effect of sunlight, "She had a frowny look, but I am not sure whether her fatigue or the sunbeam affected her."

There was a comment that the golfer's character and cultural background affect the participant's judgment. "(In the training set,) I found that there is no pattern in each group. I think there exist different feelings on each individual (P14)."

**Table 2: Results of In-depth Interview**

| Features | Type | Number of Speakers | Prediction for the next day | Main Comments |
| --- | --- | --- | --- | --- |
| Eyes | Facial landmark | 7 | Negative | "(Golfers) have teary eyes in common." |
|  |  |  | Negative | "You can read her eyes inordinately quaver." |
|  |  |  | Negative | "(Golfers) have trembling eye sights." |
| Confidence | Emotion | 5 | Positive | "She could not hide to surge up her confidence." |
|  |  |  | Negative | "Excited-states ends up the bad score, but it is hard to tell apart from pride." |
|  |  |  |  | "Her emotion looked different from |
| Gesture | Pose | 5 | Negative | "With her arms crossed, I felt a passive attitude." |
|  |  |  | Negative | "The small gestures meant defensive mode for me." |
|  |  |  | Negative | "A larger gesture and motion seemed; she wanted to justify her unexpectedly nice strokes of the day." |
| Lips | Facial landmark | 3 | Positive | "I read the speed of the lips. Speeding up means, she is pretty excited about her performance." |
|  |  |  | Negative | "I read exhausted image on her… her lips motioned very fast." |
|  |  |  | Negative | "… while they listen to the interviewer, keep their mouse opened." |

| Features | Type | Number of Speakers | Prediction for the next day | Main Comments |
|---|---|---|---|---|
| Fatigue | Emotion | 3 | Negative | "(she looked like) she wanted to go back to the room and took a shower." |
| Sunshine | Weather (external) | 1 | Not Sure | "She had a frowny look, but I am not sure whether her fatigue or the sunbeam affected her." |
| Score | Subtitle (external) | 1 | Depends on score | "She shot nine-under on that day, so I thought she would not make such a perfect score on the next day, too." |

*4.3.3* **Promising features from humans**

Both participants and quizzes are too small to generalize; however, we found some interesting differences between correct answerers and incorrect answerers, which could be further explored in future work. Based on the interview contents of 100% correct and 100% incorrect answerers, the former mentioned much more details about the golfer's face. Unlike the latter, which focused more on non-verbal languages, such as gestures, correct answerers mentioned more about golfers' eyes and lips. P11 stressed, "I am the one that exactly matches the expression 'the eyes have it'", "In my case, I closely observe the lip, then figure out her confidence." P16 also emphasized, "My criteria on a guess relies on the size of eyes and the movements of pupils." Most of the correct answerers were not very knowledgeable about golf, but actively used their intrinsic insights for the quiz.

Interestingly, 100% correct answerers' jobs relate to observing and facing people (such as speech instructor, oral-test instructor, PR team manager (former journalist), and exhibition manager). They mentioned that their inherent views on people led to the correct answer. P15, who teaches oral tests to students, said, "There is something in common right before the interview starts. If video played from the middle of the interview, I might not hit the mark," indicating that he mostly focuses on golfers' default mode and just finished their round, and thus, possibly missing the part when golfers start disguising for the interview.

In addition, there exists a difference in time for viewing test videos. Half of the 100% correct answerers turned off the video for less than 1 minute. Also, among the 50% correct answerers, two of the six watched the video shortly and made a good guess. However, even if answered in a short time, their certainty on themselves was not that high.

We conducted the study in February, 2020. Due to the rapid spread of COVID-19 in South Korea, we could not meet more people. As of now, schools and offices have been still locked down. Moreover, in South Korea, golf is a popular sport. Thus, people are quite knowledgeable about golf players and rules. In addition, they might have a bias in reading foreigners' faces or expressions. The side of constructionism criticized that juries in a single culture cannot guarantee judgment by reading the face of the suspect [25]. To overcome these limitations, we also employed meta-data into the algorithm.

## 5 Study 2: micro-expressions + Meta-data analysis

In this study, we put human selected features from the Section 4 into machine learning process, which represents the third(III)) and the fourth(IV) quadrant in our theoretical framework. We will compare this combination model with the human selection model and the machine's meta-data model, respectively.

Before the machine analysis, we should discuss the perception that varies in humans and machines. Zhang et al. asserted that machines use more contextual information than humans for object detection in an image. While recognizing kimono in the picture, segments like Oshiroi makeup and hairstyle around kimono were selected to machine's decision-making grounds, while humans only focused on costume [76]. For using patterns from image itself, however, we need a large dataset. We used cues from humans and meta-data to imitate the machine's pattern recognition with a tiny-dataset.

## 5.1 Input Data

In this section, we subjected the 213 videos mentioned in Section 4.1 to machine analysis. The size of the dataset is much smaller than that used in few-shot learning in general. We use time-series frames in these videos, where golfers mostly show frontal faces.

Table 3: Variables of the dataset

| Columns for facial features (8 ×100) | Columns for meta-data |
| --- | --- |
| Left_upper_eyeline_0~99, Left_lower_eyeline_0~99, Right_upper_eyeline_0~99, Right_lower_eyeline_0~99, Left eyebrow_0~99, Right eyebrow_0~99, Mid of lip_0~99, Right_end_of_lip_0~99 | Age, Length of career, Height, Rank of the previous tournament, Nationality (one-hot encoded) |

The balance of the dataset was 0.56:4.25; the video amount corresponding to the decrease in the score were much more than the opposite. We applied weights for each class in Experiment 2, the deep-learning analysis for making up for the imbalance.

In the experiment, we employed features selected by correct answerers from Section 4.3. We used up to 100 frames for videos, with a maximum of over 900 frames. However, most of the correct answerers tended to watch videos for less than one minute. Then, we tracked the golfer's face in frames, extract landmarks, especially eyes and lips. We included professional years, age, nationality, height, and rank of the previous tournament, as shown in Table 3. The reasons for selecting those meta-data can be described as follows.

First, in our ground theory, the rank of the previous tournament crucially affects the next tournament. Tiger Woods, the winner of the 2019 Masters at Augusta National Golf Club, said, "The win at East Lake was a big confidence-booster for me [4]." The PGA Tour Championship at East Lake that Woods mentioned was not just before the Masters; however, the previous experience affected his rebound. "East Lake was a big step for me confirming that I could still win out here and against the best players", he said [59]. We believe that matches in the past could affect golfers psychologically.

Length of career and ages were used to measure competence in interviewing. We often listen to comments like "(to the teenage golfer) debuted as professional recently, she cannot manage her expressions and hide emotions, because she is too young (10 September 2020, at ANA inspiration, comments by sportscaster in South Korea)." We believe the number of interviews affects facial movements in front of the camera, so we included the gap between the interview year and debut year into the feature.

We also considered nationalities. In the LPGA tour, golfers from various countries play events. In 2008, the LPGA commissioner announced a requirement to speak English sufficiently to all golfers on tour. The comment sparked a backlash, then swiftly being backtracked [13]. For non-native speakers, the English interview could be a burden; therefore, the facial expressions may vary depending on nationality. Moreover, as mentioned in Section 2, emotions could correspond to different feelings in different cultures. Meanwhile, we included the Golfer height in our feature set as it represents the physical state of the athlete.

## 5.2 Method

To evaluate time-series movements in facial landmarks, we employed long short-term memory (LSTM). For the embedding factors of our prediction algorithm, the changes in facial coordinates and meta-data features were merged. Other features discovered in videos, such as the number of galleries, weather, and subtitles informing scores, were omitted. We tried to focus on the psychological state and future score of individual golfers, and thus, excluded external factors from the video.

We split the video data into frames based on 0.25-second time range. Generally, in video studies, researchers cut frames based on action units (AU). However, in the interview scenes, few dynamic changes were observed. In this study, as faces with delicate changes in time series were observed, we decided to split them apart as short as possible. Commonly, frames by 0.25 seconds are widely used in computer vision studies; we have empirically identified the number that does not damage the face to any video transition and verified that 0.25 seconds is enough for our purpose. Even with precise design, the python-based OpenCV library failed to correctly detect the face in some frames, resulting in several missing frames. Further studies are needed to explore a better method to reduce the loss.

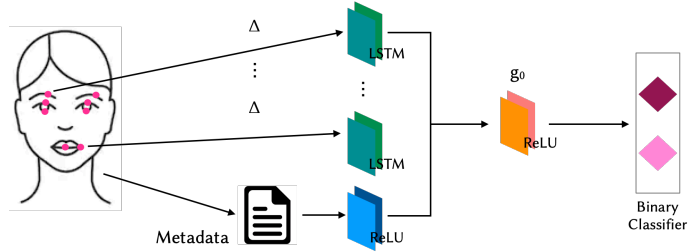

**Figure 4: Structure of the algorithm. Derived from facial landmark coordination, we calculate deltas from each frame. We drove each facial landmark delta into the lstm layer, then merged all outputs, including meta-data output, trained by fully-connected networks, into the g0 layer. Through g0, then the final results, determining the player's next day score, comes out.**

The structure of the machine learning algorithm is described in Figure 4. We collected coordinates of facial landmarks, including eyes and lips, using the python-based Dlib library (dlib.net). This library makes easier to get specific coordinates from faces. We used eight x-coordinate points of the face, in total, described in Table 3 and Figure 4 (with pink points). Then, we listed the numbers frame by frame and converted them into movement values (delta). Depending on the frame size, the delta could vary among videos. Thus, we modified all deltas into relative numbers. To avoid the overfitting issue, we employed dropouts on every layer. We set $f_0$–$f_8$ as a LSTM layer, with hidden dimensions 10. For $f_9$, to deal with meta-data, we also constructed a two-layer fully connected network, with hidden dimensions 128 and 64, respectively, with ReLU as their activation function. The output dimension of $g_0$, after merging $f_0$–$f_9$, is a binary, with ReLU as its activation function.

For evaluation, we juxtaposed three tests, putting differences on input features, using deltas of facial landmarks only, meta-data only, and merged features with deltas and meta-data. We used the F1-score for comparison as the dataset was imbalanced. The F1-score for a single class is given by the harmonic mean of precision and recall, as described in Equation (1):

$$F1 = 2\frac{Precision \times Recall}{Precision + Recall} \qquad (1)$$

## 5.3 Results

These tests revealed that the algorithm merging with the meta-data and the selected features by people performed better than that using features separately. The merging model obtained an F1-score of 0.6047 on test-set, which was higher than those obtained solely by using facial features (0.3488), or meta-data (0.4826). There exists a tendency for algorithms with more features to achieve better performance in general. However, in small datasets with imbalanced classes, it demonstrates inverse performance. Therefore, this result is significant at a highly skewed and tiny-data level.

# 6  discussion

In summary, we employed human insight into the machine learning process to overcome small-dataset limitations. We figured out the people who gained insights on knowing others and gathered their selected features. We designed an experimental model to predict the future score by watching the LPGA golfer's interview video. From the video, people chose facial landmark movements as important indicators for their prediction. We input people's ideas and other meta-data, such as previous scores and nationalities, into the algorithm. As a result, merging these two kinds of features resulted in a better F1-score than using features separately. In this section, we discuss the implications and limitations based on the obtained results.

### *6.1.1* Phenomenological insights into human-in-the-loop AI model

An important goal for AI research today is to employ deep learning based approaches on small datasets without overfitting issues. As a theoretical implication, we proposed insightful participation in the deep learning process to solve the problem. Meanwhile, as HCI researchers, we linked our model to the human-in-the-loop AI model regarding people's focusing areas being actively engaged in the modeling process of machines. Merging human ideas and machine perception led to better performance. We will further explore better human-AI collaborative work for humane factor analysis.

Furthermore, our contextual framework – brought the idea from the phenomenology – showed the possibility to drive the human-in-the-loop AI model abundant. We employed phenomenology of intersubjectivity context to perceive other's minds. In this theory, we can directly perceive others' minds, not only by relying on primary factual data from their bodies but also by using some contextual frameworks for interpretation based on the experiences of a life shared by us and them. Reconsidering the human as an embodied neuropile of an algorithm, humans' perception of others' changes in the face combined with her/his meta-data could contribute to the neural network framework, as we reported above. Then, we could use this algorithmic decision to judge other's state in practical uses, inversely.

For another practical implications, our method was designed to minimize the harmful effects on facial recognition and relevant analysis that recently arouse [9][75]. There exist two criticisms regarding facial recognition algorithms: potential violation of personal privacy, and possible discrimination against some groups due to the embedded algorithmic bias learned by humans. The former is believed to be managed by legal and institutional frameworks to a certain extent. However, solving the latter is quite challenging. Therefore, IBM [52] declared that they would no longer develop facial recognition solutions. Microsoft and Amazon also announced that they would not provide their relevant algorithm service to law enforcement institutions in the U.S. [7][67].

We intentionally avoided using emotional labels to overcome cultural biases, especially for emotion recognition. Instead, we focused on whether patterns made by facial movements and players' meta-data are related to performance. Consequently, our contribution relies on the analysis of sports data to reduce potential algorithmic bias.

### *6.1.2* Limitations due to COVID-19

This study suffered from some limitations. Due to the COVID-19, our informants were all recruited from our country, as such, they were quite knowledgeable to golf game and golfers and their views may not be fully representative of the broader class of the population. Also, there could be a dispute on the small number of questions, which cannot guarantee the correct answerers as being an 'insightful people'. In future work, we should compare the results of this study with the model with an incorrect answerer's idea being applied, using more quizzes. In addition, the feature-selection on meta-data relied too much on the hypothesis based on our ground truth. A more objective process for features, such as the analysis of ranks of previous events related to future scores, is needed.

Notably, the audio was excluded from this psychological analysis. As mentioned above, golfers mostly speak about similar contents such as "I will (did) just concentrate on my golf," and there exist more and more variables to English fluency and personal variables about voice and vocal style. Generally, an increase in variables could lead to overfitting in a small dataset analysis. Thus, we excluded audio-related features. In our future work, we will investigate multimodal methods with more data.

Moreover, we will build our method using the more generalizable human selected model in other types of tiny-datasets, and other games such as shooting or archery. We hope that this method could collaborate with on-site staff and experts in sports to elaborate on the psychological analysis. Furthermore, education in COVID-19 could benefit from a more delicate model. By collecting facial appearances of students, not only the activity and concentration, but also their emotions could be analyzed by the model, where educators' ground truth could be embedded. We believe that the proposed method will help establish a more practical and intuitive human-in-the-loop model, and we are eager to see how it contributes to human-AI interaction studies.

# 7 Conclusion

We have shown that combining people's insight with a deep learning algorithm in a tiny-dataset can substantially contribute to improvement of score prediction performance using facial landmark movements and meta-data. We have specifically leveraged perception feedback from people with good insight into decoding other's minds. Insightful people have picked transitions of players' emotions for their certainty, focusing on eyes and lips movements. We also applied meta-data into our algorithm, such as players' previous score, height, nationality, age, and length of career, which reflect players' backgrounds. An integrated model demonstrated a better F1-score than other models that dealt with features separately. Based on the results, we suggest that the human-in-the-loop model, based on our phenomenological framework could improve the algorithm performance with a tiny-dataset. Future work will further explore the generalizability of other small data analyses.